\documentclass[manuscript]{aastex}

\begin{document} 

\title{Hubble Space Telescope Ultraviolet Light Curves Reveal Interesting Properties of
CC Sculptoris and  RZ Leonis} 

\author{Paula Szkody\altaffilmark{1,2},
Anjum S. Mukadam\altaffilmark{1,2},
Odette Toloza\altaffilmark{3},
Boris T. G\"ansicke\altaffilmark{3},
Zhibin Dai\altaffilmark{4,5,6},
Anna F. Pala\altaffilmark{3},
Elizabeth O. Waagen\altaffilmark{7},
Patrick Godon\altaffilmark{8},
Edward M. Sion\altaffilmark{8}}

\altaffiltext{1}{Department of Astronomy, University of Washington,
  Box 351580, Seattle, WA 98195; szkody@astro.washington.edu}
\altaffiltext{2}{Based on observations obtained with the Apache Point
  Observatory (APO) 3.5-meter telescope, which is owned and operated
  by the Astrophysical Research Consortium (ARC).}
\altaffiltext{3}{Department of Physics, University of Warwick, Coventry CV4 7AL, UK}
\altaffiltext{4}{Yunnan Observatories, Chinese Academy of Sciences, 396 Yangfangwang, Guandu District, Kunming, 650216, P. R. China}
\altaffiltext{5}{Key Laboratory for the Structure and Evolution of Celestial Objects, Chinese Academy of Sciences, 396 Yangfangwang, Guandu District, Kunming, 650216, P. R. China}
\altaffiltext{6}{Center for Astronomical Mega-Science, Chinese Academy of Sciences, 20A Datun Road, Chaoyang District, Beijing, 100012, P. R. China}
\altaffiltext{7}{AAVSO, 48 Bay State Rd, Cambridge MA 02138}
\altaffiltext{8}{Department of Astrophysics and Planetary Science, Villanova University, Villanova, PA 19085}
 
\begin{abstract}

Time-tag ultraviolet data obtained on the Hubble Space Telescope in 2013 
reveal interesting variability related to the white dwarf spin in
the two cataclysmic variables RZ Leo and CC Scl. RZ Leo shows a 
period at 220\,s and its harmonic at 110\,s, thus identifying it as a likely 
Intermediate Polar (IP). The spin signal is not visible in a short 
single night of ground based data in 2016, but the shorter exposures in that 
dataset indicate a possible partial eclipse. The much larger UV amplitude of the
spin signal in the known IP CC Scl allows the spin of 389\,s, previously only 
seen at outburst, to be visible at quiescence. Spectra created from the peaks 
and troughs of the spin times indicate a hotter temperature of several
thousand degrees during the peak phases, with multiple components contributing
to the UV light.

\end{abstract}

\section{Introduction}

The advantage of spectral data obtained in time-tag mode with the Hubble Space Telescope 
(HST) is that light curves can be constructed by summing the flux over wavelength.
Thus, we were able to search for periodic
ultraviolet variability in the observations resulting from a large Cycle 20 $\it{HST}$ program with the Cosmic Origins Spectrograph (COS) 
that was designed to obtain the temperature of the
white dwarfs in 40 cataclysmic variables (CVs).
In CVs with non-magnetic white dwarfs and with short orbital periods
(thus low rates of mass transfer), the accretion disk typically contributes 
40-75\% of the optical light 
whereas the white dwarf contributes 75-90\% of the ultraviolet (UV) light 
(Szkody et al. 
2010). Of the 40 systems in our $\it{HST}$ program (Pala et al. 2017), 
two showed periods
identifying them as new non-radial pulsators (Mukadam et al. 2017) and five others
had observed eclipses (Pala et al. in prep). Here we report interesting results
on the UV variability of two systems that are related to the spin of their
white dwarfs: in RZ Leo, we detect for the first time a probable spin period of the white dwarf
while CC Scl shows a spin period at quiescence that was previously only 
evident at outburst.

RZ Leo is a short period (1.825 hr) dwarf nova with a quiescent V mag of 18.8 
and 7 past known outbursts (Mennickent
\& Tappert 2001, Ishioka et al. 2001, Patterson et al. 2003). Its optical 
spectrum at quiescence (Szkody et al. 2003) shows
broad absorption surrounding double-peaked Balmer emission lines, 
indicative of a high inclination, low mass transfer system. A short 
cadence (1 min) observation sequence
by Kepler K2-1 for 82 days from 2014 May 30 to August 20 (Dai et al. 2016)
greatly improved the orbital period determination and showed a 0.5 mag 
double-humped light variation phased on this period. During the
K2 observation, it also showed an unusual brightening event of about 0.6 mag
that lasted almost 2 hrs.

CC Scl is a $\it{ROSAT}$ X-ray source that was identified as a 17.3 mag CV by 
Schwope et al. (2000).
Further followup observations revealed spectra with strong Balmer and HeII 
lines, short duration (9 day) outbursts with superhumps, 
and an orbital period of 1.40 h (Chen et al. 2001, Ishioka 
et al. 2001, Woudt et al. 2012). Kato et al. (2015) later used shallow
eclipses to determine an inclination of 81 deg, and refine the period to 84.337 
min. Using the 2014 superoutburst data, they found a low mass ratio 
q= 0.072$\pm$0.003. Longa-Pena et al. (2015) confirmed
this low value of q from spectroscopy and suggested that CC Scl was a 
post-period bounce system. Observations with $\it{Swift}$ along with 
ground-based photometry obtained during an outburst in 2011 (Woudt et al. 2012)
revealed a white dwarf spin period of 389.5\,s which was only visible during the
outburst. The presence of this spin period identified CC Scl as a member of
the intermediate polar (IP) 
group\footnote{http://asd.gsfc.nasa.gov/Koji.Mukai/iphome/catalog/alpha.html}, 
systems containing magnetic white dwarf whose spins 
are not synchronised with their orbits; see Warner (1995) for a review of IPs.
Superhumps with a period of 1.443 h were also seen during this outburst,
making this system, along with V455 And (Araujo-Betancor et al. 2005a), 
unambiguous superhumpers containing a magnetic white dwarf.
 
\section{Observations}

The Cosmic Origins Spectrograph (COS) was used with the G140L grating to obtain
time-tag spectra for the two objects in 2013. A general description of the 
program for temperature determination is given in Pala et al. (2017). The 
number of $\it{HST}$ orbits used 
for each system varied based on their quiescent magnitude.  Four $\it{HST}$ 
orbits 
were used for RZ Leo on April 11, and two for CC Scl on June 29. The data were 
obtained
from the archive and spectra were extracted using a 41 pixel width. 
For period analysis, light 
curves were created by summing the spectra over a region with good signal 
(but leaving out strong emission lines) and binning the flux into 5 s bins. The times and spectral
regions are summarized in Table 1. The
resulting light curves were then divided by the mean and one was subtracted 
to produce a fractional amplitude. A Discrete Fourier Transform (DFT) period 
analysis was then
accomplished, with the 3$\sigma$ noise level determined by a shuffling 
technique (see Szkody et al. 2012 for details).

Monitoring of each system through AAVSO alerts and subsequent 
observations prior to the $\it{HST}$
scheduled dates determined that each system was at quiescence.
Kepler K2 continuous short cadence 1 minute observations on RZ Leo took 
place in 2014 from  May 30 to August 20.
The photometric extractions, resulting light curve, and a
discussion of the observed double-humped orbital period variations are 
described in Dai et al. (2016). In this paper, we subjected the short cadence (1 min) 
dataset to the same period analysis procedure as for our $\it{HST}$ data.  

Further optical photometric observations of RZ leo were accomplished at
Apache Point Observatory on 2016 May 29 using the 3.5m telescope with the 
frame-transfer CCD Agile (Mukadam et al. 2011) and a BG40 broad band filter. 
Integration times were 15 sec and differential light curves were constucted
using comparison stars on the same frames. The light curves  were then
converted to fractional amplitudes and the DFT computed 
in the same way as for the $\it{HST}$ data. 

\section{Results}

\subsection{RZ Leo}

The UV light curve for the four $\it{HST}$ orbits and the resulting DFT are
shown in Figure 1. The double-humped orbital variation at 55 min that has been 
observed in the optical (Mennickent et al. 1999, Patterson et al. 2003, Dai et 
al. 2016) is also evident in the UV light
curves (the middle of orbit 2 and the beginning and ends of orbits 3 and 4) and
in the DFT. The amplitude of this variation is slightly 
less in the UV (0.3-0.4 in
magnitude units) compared to 0.5 mag (Dai et al. 2016) in the optical, although
Mennickent et al. (1999) note a large optical variation of the humps during
their 11 year study. The model
fitting of the UV spectrum by Pala et al. (2017) determined the white 
dwarf has a temperature of 15,014$\pm$638 K and contributes 83\% of the
UV light. The larger amplitude of the orbital modulation in the optical 
would be consistent with an origin associated 
with the accretion disk. While a double-humped orbital
variation is fairly common among short orbital period systems, (e.g. WZ Sge),
 the cause
of this variation has been ascribed to several sources. Osaki \& Meyer
(2002) relate it to a spiral arm structure in the disk when the disk
has reached the 2:1 resonance. Skidmore et al. (2002) found that infrared
photometry was consistent with changing views of a hot spot during an orbit,
while Mennickent et al. (1999) invoked moving hot spots to account for the
changes in the humps over time in RZ Leo. 
In addition to the double-humps, the 
systems SDSS J080434.20+510349.2 and SDSS J123813.63-033933.0 share the common
characteristic with RZ Leo of small brightenings (Aviles et al. 2010), 
although the former systems
are thought to contain brown dwarfs while RZ Leo has a longer period and 
appears to have a normal
main sequence secondary (Mennickent et al. 1999).
 
The DFT for RZ Leo also shows significant high amplitude signals at 220 s and its harmonic of 110\,s,
with the amplitude of the harmonic (40 millimodulation amplitude; mma) being 
slightly larger than the 220 s (32 mma).
Lower amplitude periods of 206\,s and 106\,s also appear close to these periods.
The beat period of the close periods of 220\,s and 206\,s would be
41.7 min, which does not correspond to any observed period.
The long duration (80 days) of the short cadence K2 optical 
data in 2014 is ideal for searching for small amplitude periodic signals.
The result reveals a
period at 220.65\,s with an amplitude of about 6 mma, along
with linear combination periods (196\,s) with the orbital period of RZ Leo and 
harmonics (Figure 2). The one
minute cadence of K2 was too long to pick up the presence of the 
110\,s harmonic, or the close 206\,and 106\,s 
periods. Unfortunately, these periods of 220 and 196\,s are known artifacts in
short cadence K2 data (the
8th and 9th harmonics of the long cadence 29.42 min sampling of K2, see 
Gilliland et al. 2010). Thus, the K2 data cannot confirm the optical presence 
of the periods seen in the UV observation. Possible interpretations of the
UV periods include white dwarf pulsation or spin. Accreting white dwarf
pulsators generally do not show harmonics of the pulse in the UV and the 
pulsations wander slightly in frequency over time (Szkody et al. 2010). The spin
periods of white dwarfs are very stable over time and usually show up in
X-ray observations. 
A few IPs (e.g. CC Scl, V455 And) 
show both the spin and the harmonic of the spin period, with the harmonic 
stronger than the rotation period at times (Woudt et al. 2012, Mukadam et al. 
2016).  
Reis et al. (2013) reported $\it{Swift-XRT}$ data and modeling of RZ Leo, 
determining a 0.5-10 keV luminosity of 
7.4$\pm$0.4 $\times$ 10$^{29}$ erg s$^{-1}$, (10 times the luminosity of
V455 And). This is on
the low end, but within the range, of IPs. The observation was too
short (3798 s) to determine short timescale periods.   
The UV presence and strength of this
period of 220\,s and its harmonic, implies that this is the spin period of
the white dwarf and RZ Leo is likely an IP, but longer X-ray and optical
observations will be needed to confirm this identification.

The single night (4 hrs) of APO data (Figure 3) does not show the 220\,s period.
While the noise level in the DFT is much larger than the K2 data due to the
short length of the dataset, the increased aperture and time resolution results in a 3$\sigma$ noise
level of 6.5 mma. However, the 2016 APO light curve looks different than the 
2014 K2 one,
in that the double humps are of similar amplitude, whereas the typical
RZ Leo light curve has one hump with about half the amplitude of the other. 
It is possible that different accretion levels 
may affect the ability to detect the spin in the 
optical.
  
The short exposures of the APO data (20 s), compared to the 1 min cadence
of K2 and the 4-5 min exposures of Patterson et al. (2003) and Mennickent
et al. (1999), reveal possible 
eclipse-like features near times of 4800\,s and 11,400\,s (in Figure 3), which
correspond to the orbital period. The short duration
of this feature (about 3 min) could account for why it was not evident
in previous reported data. Phasing and folding our data with the ephemeris
in Dai et al. (2016), based on the maximum of the primary orbital hump,
shows the repeatability of the eclipse shape (Figure 4).
However, further photometry with short exposures and
over several days will be needed to confirm if this is indeed a partial
eclipse and if the spin period can be detected in the 
optical.

\subsection{CC Scl}

In contrast to RZ Leo, the UV light curve of CC Scl (Figure 5 top) is 
completely dominated by the
spin modulation of the white dwarf and its harmonic that Woudt et al. (2012)
only saw during the 2011 November outburst. 
The DFT (Figure 5 bottom) shows the 
harmonic of the spin at 194.6\,s with a much higher amplitude (220 mma) than 
the spin period of 389\,s (90 mma). Our UV data show that the spin modulation 
is clearly present at quiescence. Since the Woudt et al. (2012) observations 
were accomplished using the $\it{Swift}$ UVOT with a filter centered at 2246\AA, it is possible
that the variability at longer UV wavelengths has lower amplitude 
due to a high temperature
of the accreting areas. Pala et al. (2017) determined a white dwarf temperature 
of 16,855$\pm$801 K for CC Scl from the average spectrum, but note that the 
white dwarf only contributes about
35\% of the total UV flux (using a power law or constant flux additon for the
remainder) and so the temperature is not very reliable. At 
outburst, the X-ray and UVOT data of Woudt et al. (2012) 
show only the primary spin
period and not the harmonic (although the optical shows the harmonic at
varying amplitudes). As the outburst faded over 8 days, the amplitude of the 
spin modulation decreased until it was invisible at 9 days past outburst peak. 
This led Woudt et al. (2012) to postulate that the
higher accretion rate at outburst resulted in the disk blocking the second pole,
while at quiescence both poles are seen but the second pole is anti-phased with
the primary so the modulation is cancelled out. Our $\it{HST}$ observation 
disproves this idea and indicates that both poles contribute to the spin 
modulation without cancellation.
The optical data taken for CC Scl around the time of our $\it{HST}$ observation
(Pala et al. 2017) show a small outburst (magnitude 15.3 versus the usual
13.4, and duration only 5 days) on 2013 June 13, 16 days prior to the 
$\it{HST}$ data, and a return to quiescence on June 16. While it is possible 
that this event triggered a longer visibility of the spin modulation, we 
consider this unlikely due to the smaller, shorter outburst in 2013 versus 2011, and
the longer period in quiescence preceeding our observation.

In order to estimate the temperature of the accretion 
areas, we applied a procedure similar to that used to study the temperature 
variations in 
the dwarf nova GW Lib (Toloza et al. 2016). This involved using
 the Markov chain Monte Carlo (MCMC) 
ensemble sampler to fit two spectra, one obtained from the peaks of 
the lightcurve and a second obtained from the troughs. The count rates used
to create the peak and trough spectra for CC Scl are shown by the lines in 
Figure 6 and the resulting two spectra  
are shown in Figure 7. Besides the flux difference, the broader Ly$\alpha$
absorption line and the flatter
shape of the trough spectrum are indicative of a temperature decrease.  

We employed a grid of white dwarfs models (Hubeny \& Lanz 1995) 
covering a temperature range of 
9000-30000\,K in steps of 100\,K with the metallicity set to 0.2 times 
the solar abundance and with $\log{g}$=8.35 (Pala et al. 2017), corresponding
to the average mass of white dwarfs in CVs (Zorotovic et al. 2011).
The airglow emission lines of Ly$\alpha$ and \ion{O}{1} were 
masked out in the range of 1207.20-1225.26\,\AA\ and 1295.30-1312.44\,\AA, 
respectively. The emission lines of \ion{C}{3}1176, \ion{C}{2}1335, 
\ion{Si}{4}1400, and \ion{C}{4}1400 were fitted with gaussians.

The core of the Ly$\alpha$ absorption reveals evidence for a second component. 
This second component has been identified in many dwarf novae 
(Godon et al. 2004, Long et al. 2009, Sion et al. 2003, G\"ansicke et al. 2005),
although its nature and origin remain unclear. In dwarf novae, this component
is likely related to the disk and/or boundary layer emission. For IPs,
the situation is further complicated by additional  contributions to
the UV light from the accretion curtains, and the heated white
dwarf areas at the magnetic poles. The UV observations of many polars
show large variations of the white dwarf temperature during their orbits
as large areas near their magnetic poles that are 
heated by irradiation from the accretion columns come into view. This
has been studied in most detail in the prototype AM Her
(G\"ansicke et al. 1995, 1998, 2006; K\"onig et al. 2006), but
is a general feature of this class of strongly magnetic CVs (Stockman et al.
1994, Schwope et al. 2002). Even the treatment of this heated spot area is
problematic as a realistic model needs to encompass a
 temperature change from the
center of the spot to the unheated white dwarf (G\"ansicke et al. 2006) as
well as non-circular spots (Linnell et al. 2010).
Due to the limited wavelength and orbital coverage of our data,
we merely approximated the second component as a power law (with its
contribution to the peak and trough spectra determined by the core flux of 
Ly$\alpha$) and then 
tried to fit the two spectra, first trying a single white dwarf, and then a 
dual temperature white dwarf with one temperature fixed to the trough
white dwarf tempearure. Even these simple models involve 4 free parameters for 
the single white dwarf (the white dwarf temperature, scaling factor, power law
constant and exponent) while the dual temperature white dwarf invokes
an additional parameter.

The shape of the light curve of CC Scl 
imitates a sinusoidal-like pattern, suggesting that the hot areas are not 
fully self-eclipsed, thus preventing a reliable estimate of the underlying
unheated white dwarf temperature. Fitting the trough spectrum of CC Scl with a 
single white dwarf model along with a power law, we obtain an upper limit to 
the white dwarf temperature of 15,612$^{+139}_{-129}$\,K, which is likely an
average of a lower temperature white dwarf and some warmer heated area on the
white dwarf. This
fit is shown in Figure 7 and involves a power law that contributes 65\%
of the observed flux along with this average temperature white dwarf.
If we invoke the same fitting for the peak spectrum, we obtain a
white dwarf of 18,751$\pm{164}$\,K, with a power law contributing 60\%
(fit shown to peak flux in Figure 7).

Since the accretion areas likely have a steep temperature gradient near
the infalling material, we also tried to model the peak spectrum using
the average white dwarf temperature found from the trough spectrum (15,612\,K) 
plus a hotter spot area that is viewed during the peak phases, plus a power
law. This
resulted in a fit with the hotter spot temperature of 
24,170$^{+848}_{-586}$\,K and an area covering less than 9.5\% of 
the visible surface of the white dwarf, with the power law component
contributing 42\%. In this fit, the average flux of the cool white dwarf 
plus its warm area had
to increase by a factor of 1.5 implying an increase of its warm spot
area as well. Since we cannot disentangle the warm spot area from
the rest of the white dwarf and the total flux is dominated by the
component modelled as a power law, the temperatures are not well
constrained. We can conclude that the temperature of the area
viewed during UV peak phases is several thousand degrees hotter than
that at the trough phases. 

An analysis of the UV spectrum of several other IPs (AE Aqr, FO Aqr, EX Hya, 
PQ Gem) has been accomplished
using the Faint Object Spectrograph (Eracleous et al. 1994, de Martino et al. 
1998,1999; 
Stavroyiannopoulos et al. 1997) and the Space Telescope Imaging Spectrograph
(Belle et al. 2003). 
Almost all of these studies showed two or more components
giving rise to the UV flux. One of the components was always part of a white
dwarf, with temperature of 23,000\,K (Belle et al. 2003) for EX Hya, 
26,000\,K for
AE Aqr (Eracleous et al. 1994), 30,000\,K for
PQ Gem (Stavroyiannopoulos et al. 1997) and 36,000\,K for FO Aqr (de Martino et 
al. 1999). The second component was a much larger black body or disk or
column that was much cooler (T$<$12,000\,K). 
The temperatures derived for the white dwarf in 
CC Scl are lower
than in these three systems. This is consistent with the short orbital period
of CC Scl, its low magnetic moment (Woudt et al. 2012), and its low mass
ratio suggestive of a post-period minimum system (Longa-Pena et al. 2015).
The average white dwarf temperature in CC Scl measured from 
the UV faint phases (trough) of 15,612\,K,
is similar to the values obtained for the short-period polars V834\,Cen, BL\,Hyi
and MR\,Ser (Araujo-Betancor et al. 2005b), indicating
a similar accretion rate (Townsley \& G\"ansicke 2009). The power laws,
which increase toward shorter wavelengths, are in good agreement with
the presence of a magnetic or other hot component.
 
\section{Conclusions}

The UV light curves obtained from time-tag COS spectral data reveal several new
facets of RZ Leo and CC Scl.
The UV light curve of RZ Leo in 2013 
shows periods of 220\,s and its harmonic at 110\,s, thus making it a likely
new member of the small class of IPs with orbital periods less than 2 hrs.
Two years later, optical ground-based photometry obtained 
over a 4 hr timescale does not show
this signal, but does indicate a possible partial eclipse. Further X-ray
and optical data are needed to confirm this classification.
CC Scl at quiescence shows very prominent UV visibility of the spin period 
previously observed only at outburst, with the harmonic having a larger 
amplitude than the spin period itself.
This indicates both accretion poles contribute to the modulation at quiescence.
Fitting white dwarf plus power law models to the spectra created from the peaks versus troughs
of the spin count rates show increases in temperature of several thousand 
degrees when the accretion poles are in view. However, since the power law 
component
is a significant part of the observed UV flux, a better understanding
of the temperature gradient and contribution of each of the magnetic
heated poles, along with a distance determination, is needed to further 
constrain the temperatures and areas involved. 

In the Mukai catalog, only about 10\% of the IPs have orbital periods
below 2 hours, like RZ Leo and CC Scl. Since both
theoretical and observational arguments have been made (e.g. Lasota et al.
1995) for the presence of truncated disks in low accretion rate short
orbital period systems, it is possible that the number of IPs in this
period range is much larger. However, as this paper demonstrates, it is
difficult to detect spin periods in these systems.

\acknowledgments

PS, ASM and EMS acknowledge support from NASA grants HST GO-12870 and 
GO-13807 from the Space
Telescope Science Institute, which is operated by the Association of
Universities for Research in Astronomy, Inc., for NASA, under contract
NAS 5-26555, and from NSF grant AST-1514737. OT was funded by the CONICYT 
becas-CONICYT/Becade-Doctorado-en-el-extranjero 72140362.
The research leading to these results also received funding from
the European Research Council under the European Union's Seventh
Framework Programme (FP/2007- 2013)/ERC Grant Agreement No. 320964 (WDTracer). 
ZD acknowledges support from CAS Light of West China Program and the Science
Foundation of Yunnan Province (No. 2016FB007).

We are grateful to the dedicated 
AAVSO observers for monitoring these objects prior to the HST observations
to ensure their quiescent states. We also thank JJ Hermes for pointing
out the artifacts present in short cadence K2 data.

\clearpage
\begin{deluxetable}{lcccc}
\tablewidth{0pt}
\tablecaption{Summary of HST COS Observations}
\tablehead{
\colhead {Object} & \colhead{UT Date} & \colhead{$\lambda$ (\AA)} & \colhead{UT Time Range}
 & \colhead{Exp (s)}}
\startdata
RZ Leo & 2013 April 11 & 1122-1980 & 08:35:43-14:08:15 & 10504.736 \\
CC Scl & 2013 June 29  & 1121-1960 & 03:39:03-05:46:30 & 4667.776 \\
\enddata
\end{deluxetable}

\clearpage

\begin{figure}
\figurenum {1}
\includegraphics[width=7in]{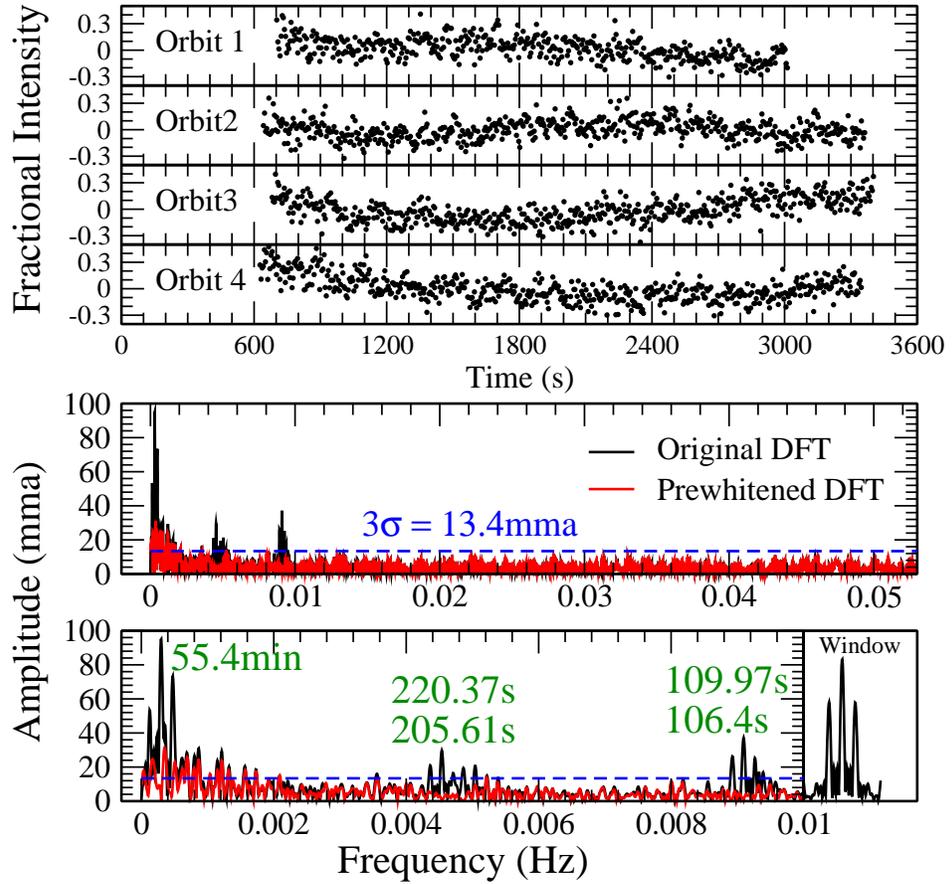}
\vspace*{-2in}
\caption{Fractional intensity light curve (top), total DFT (mid) and expanded
DFT at low frequencies (bottom)  of the 4 orbits of COS time-tag
data on RZ Leo obtained on 2013 April 11.}
\end{figure}

\clearpage

\begin{figure}
\figurenum {2}
\includegraphics[width=7in]{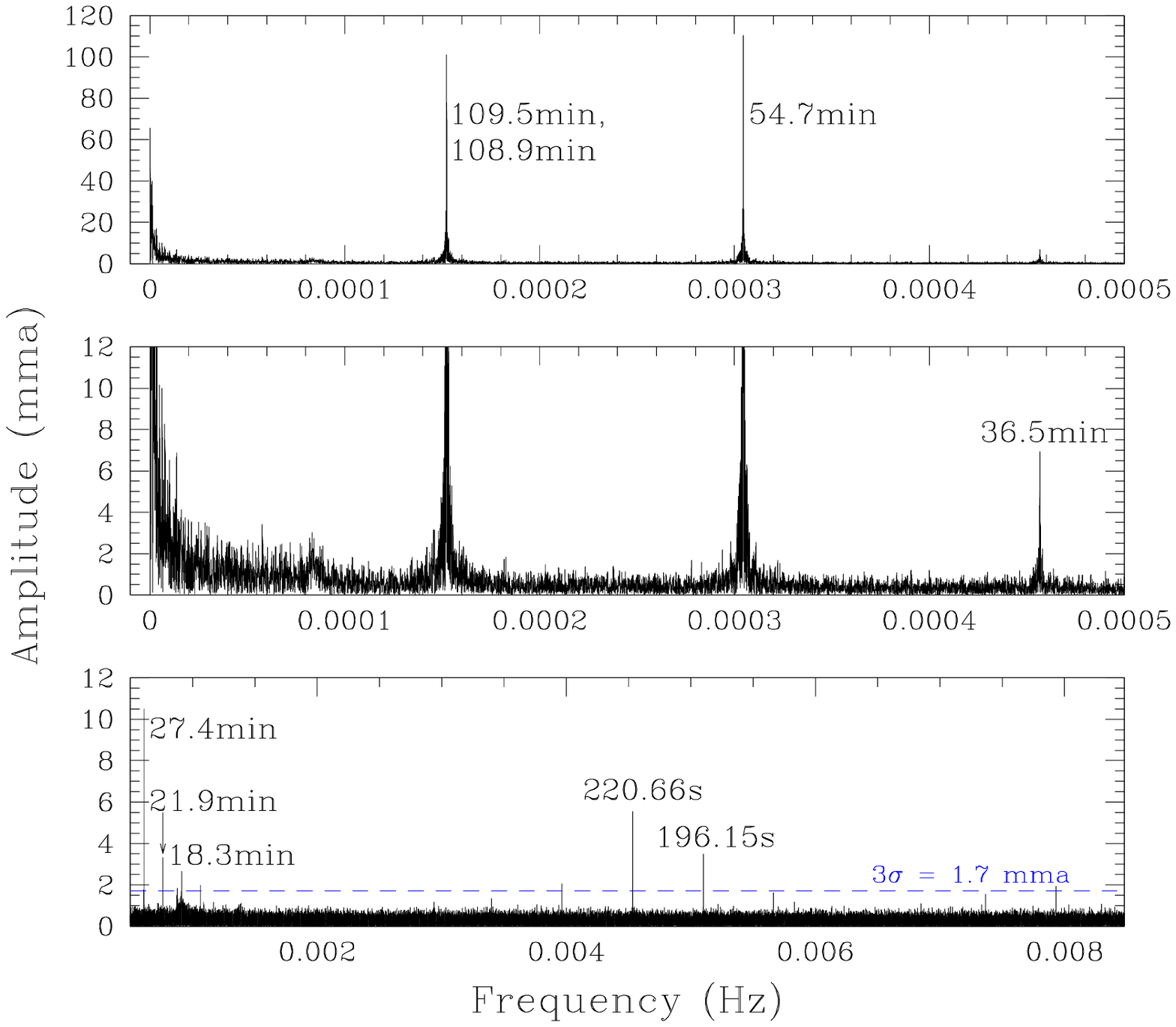}
\vspace*{-3.5in}
\caption{DFT of short cadence K2 continuous light curve data on RZ Leo from 2014 May 30 
through Aug 20. Middle panel shows the same frequency range as the top panel but expands
the low amplitudes. Bottom panel plots the entire frequency range to the Nyquist 
frequency (120 s). The peak labelled 109.5 is the orbital period, with harmonics at
54.7, 36.5, 27.4, 21.9 and 18.3 min, 220 s and 196 s are the 8th and 9th
harmonics of the K2 long cadence sampling time.
The blue line shows the 3$\sigma$ level determined  from the shuffling
technique described in Szkody et al. (2012).}
\end{figure}

\clearpage

\begin{figure}
\figurenum {3}
\includegraphics[width=7in]{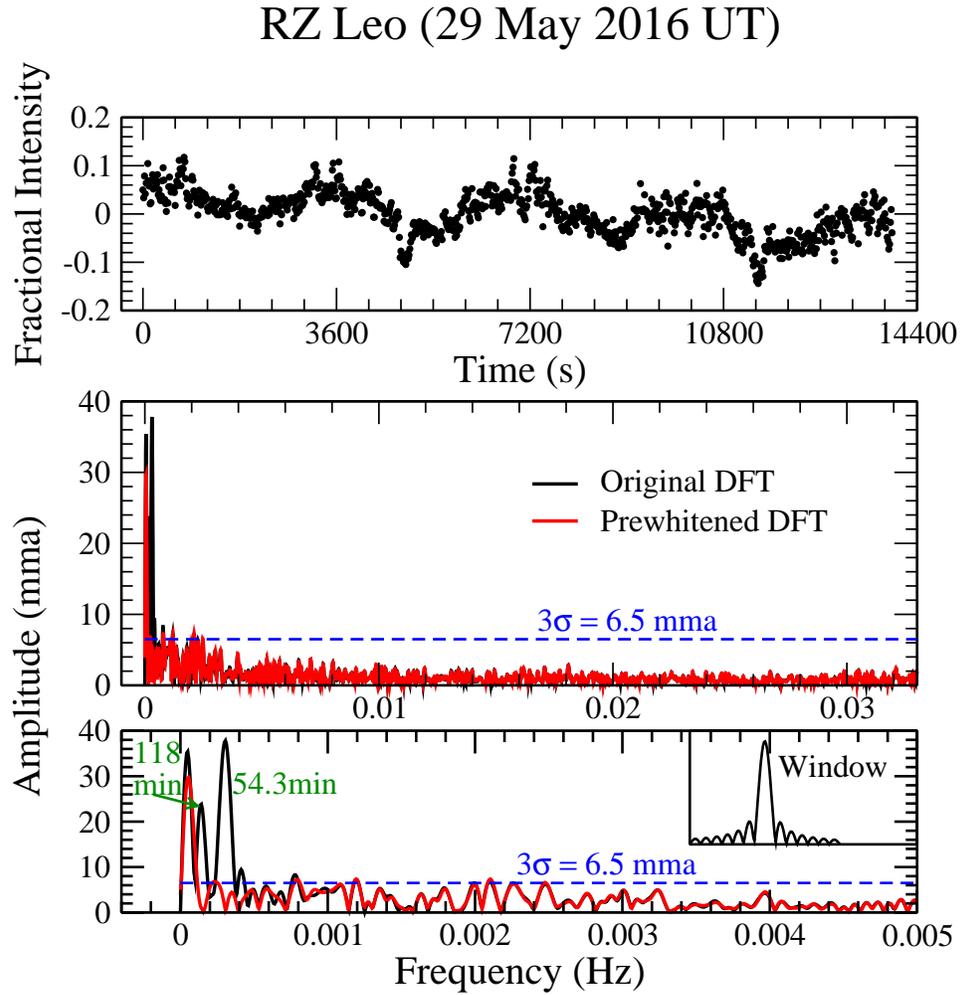}
\vspace*{-2in}
\caption{Fractional intensity light curve and DFT of RZ Leo obtained at APO on
29 May 2016 with 15 s integration times.}
\end{figure}

\clearpage

\begin{figure}
\figurenum {4}
\includegraphics[width=8in]{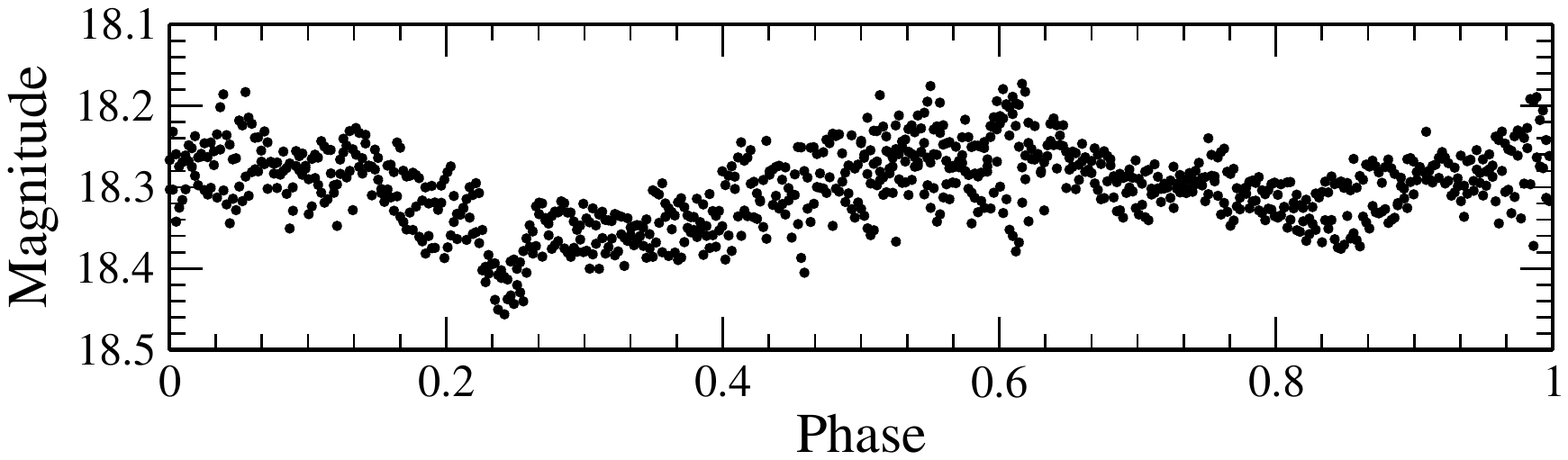}
\vspace*{-2in}
\caption{RZ Leo APO data converted to approximate g magnitudes using comparison 
stars from SDSS and phased on the ephemeris given in Dai et al. (2016), where
phase 0 is the maximum of the primary orbital hump. Each point has an error
bar of about 0.01 mag.}
\end{figure}
\clearpage 

\begin{figure}
\figurenum {5}
\includegraphics[width=7in]{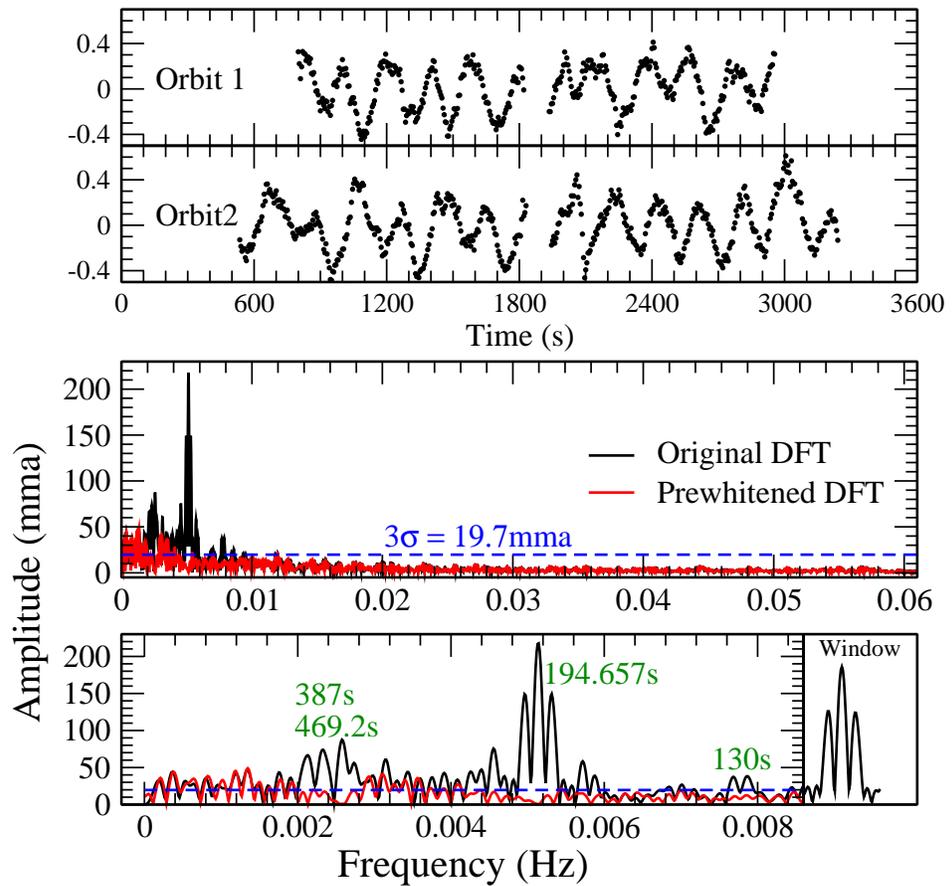}
\vspace*{-2in}
\caption{Fractional intensity light curve and DFT of the 2 orbits of COS time-tag
data on CC Scl obtained on 2013 June 29.}
\end{figure}
\clearpage

\begin{figure}
\figurenum {6}
\includegraphics[width=7in]{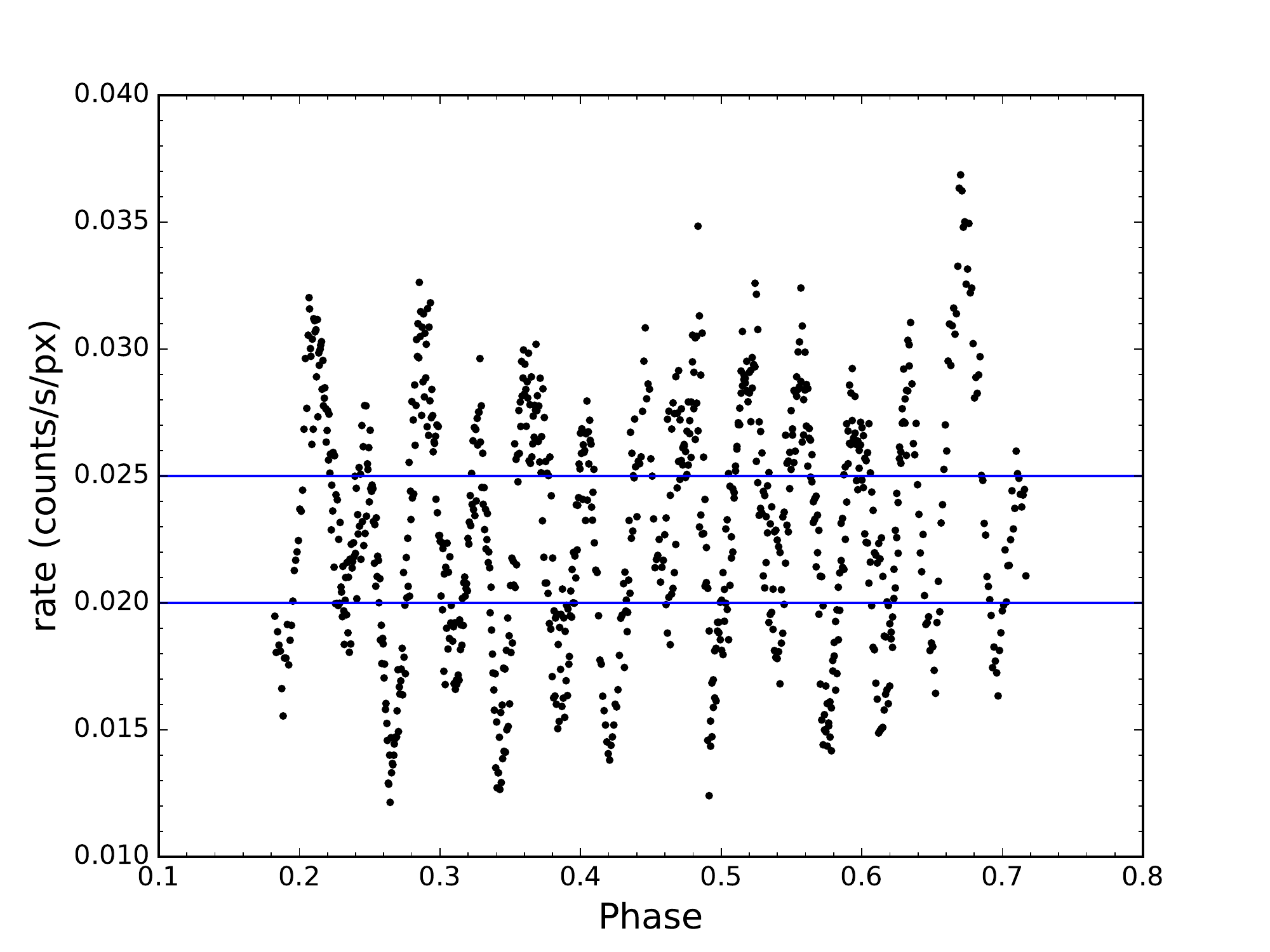}
\caption{HST Count rate light curve of CC Scl showing the regions used to create the
peak spectrum (above the top line) and the trough spectrum (below the
bottom line).}
\end{figure}

\clearpage

\begin{figure}
\figurenum {7}
\includegraphics[width=7in]{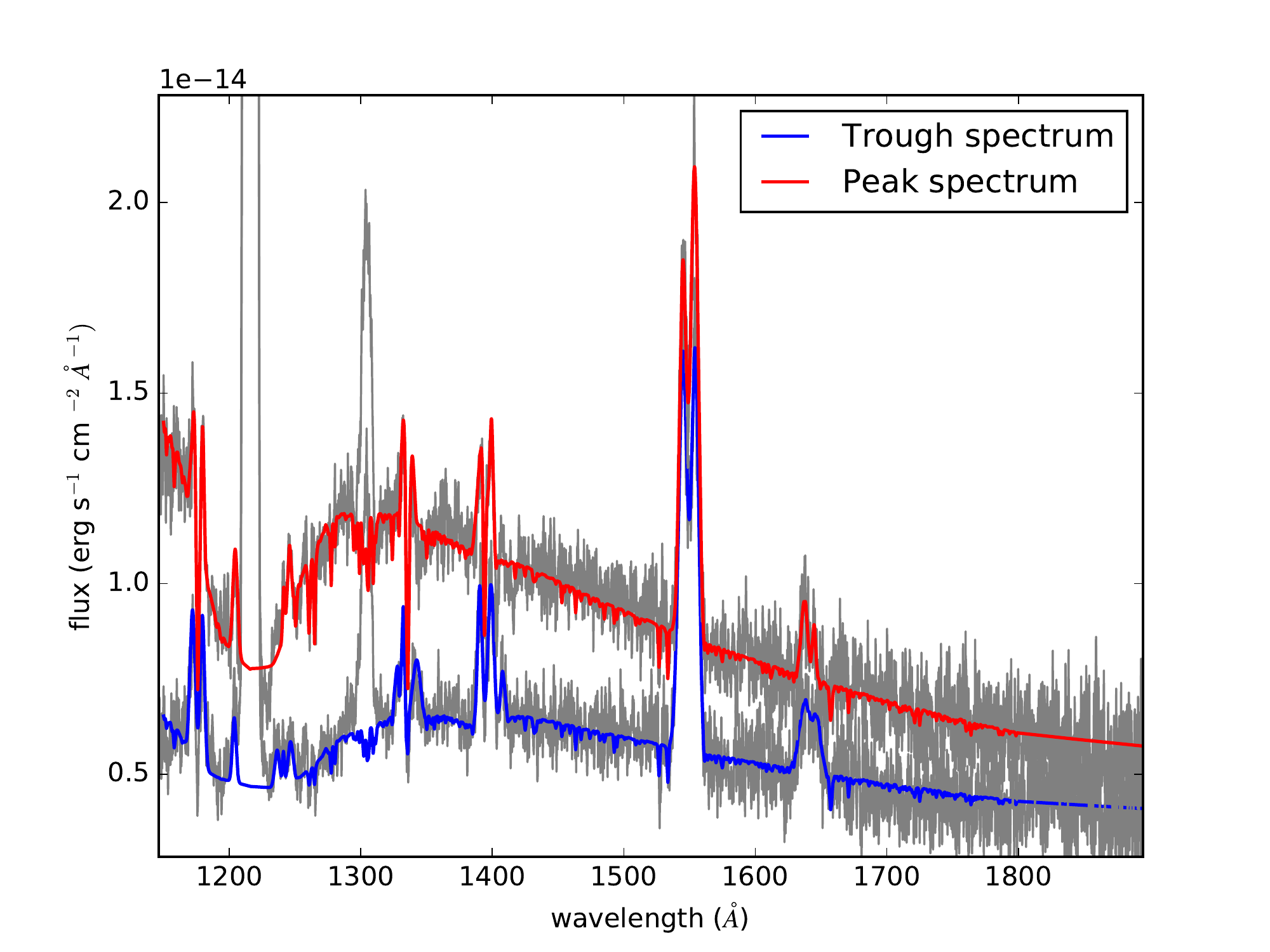}
\caption{The CC Scl 
spectra created from the peaks (top grey), and troughs (bottom grey) of the
data shown in Figure 6, along with a model white dwarf and power law fit.
The trough spectral fit (blue) is for a 15,612\,K white dwarf
along with a power law contributing 65\%, while the peak fit uses an 18751\,K 
white dwarf (red) along with a power law contributing 60\%.}
\end{figure}

\end{document}